\documentclass{kluwer}

\usepackage[dvips]{graphics,color}

\begin{document}
\begin{article}

\newcommand{\D}{\displaystyle}
\newcommand{\rsun}{\rm{R}$_{\odot}$}

\begin{opening}

\title{Phase Coherence Analysis of Solar Magnetic Activity}
\author{\surname{Carl J. Henney}}
\author{\surname{John W. Harvey}}
\institute{National Solar Observatory{\thanks{National Solar
Observatory (NSO) is operated by the Association of Universities for
Research in Astronomy (AURA, Inc.) under cooperative agreement with
the National Science Foundation (NSF).}}, Tucson, Arizona, 85726-6732, USA}

\begin{abstract}

Over 24 years of synoptic data from the NSO Kitt Peak Vacuum 
Telescope is used to investigate the coherency and source
of the 27-day (synodic) periodicity that is observed over multiple 
solar cycles in various solar-related time series. A strong 
27.03-day period signal, recently reported by \inlinecite{Neu00}, 
is clearly detected in power spectra of time series 
from integrated full-disk measurements of the magnetic flux 
in the 868.8 nm Fe~I line and the line equivalent width 
in the 1083.0 nm He~I line. Using spectral analysis of synoptic maps of
photospheric magnetic fields, in addition to constructing 
maps of the surface distribution of activity, we find 
that the origin of the 27.03-day signal is long-lived complexes 
of active regions in the northern hemisphere at a latitude of 
approximately 18 degrees. In addition, using a new time series 
analysis technique which utilizes the phase variance of a signal, 
the coherency of the 27.03-day period signal is found to be 
significant for the past two decades. However, using the 
past 120 years of the sunspot number time series, the 27.03-day 
period signal is found to be a short-lived, no longer than 
two 11-year solar cycles, quasi-stationary signal. 
\end{abstract}

\end{opening}

\section{Introduction}

The variability of solar magnetic activity, along with differential rotation 
and meridional flow, leads to a spatially and temporally dynamic solar 
surface. With such continuous change, the temporal coherence of observed 
solar magnetic activity is expected to decrease for intervals much 
greater than the average lifetime of an active region. Thus one might
expect only a broad peak covering the range of rotation periods in
spectral analysis of solar activity time series. However, the persistence 
of specific 27-day periodicities (all periods stated in this paper 
are synodic) over several solar 
cycles found in various solar related time series has 
led to numerous papers supporting the existence of preferred longitudes 
of solar magnetic activity (e.g. \opencite{Sval75}; \opencite{Bog82}; 
\opencite{Bal83}; \opencite{Neu00}). \inlinecite{Sval75} reported a 
periodicity near 27-days that lasted over five solar cycles using 
autocorrelation analysis of interplanetary magnetic field data. 
\inlinecite{Bog82} revealed a persistent $27.5 \pm 0.5$~day period signal 
using spectral and autocorrelation analysis with 128 years of the sunspot 
number time series. In addition, correlating sunspot group longitudinal 
positions with high speed solar wind, \inlinecite{Bal83} also 
found a persistent periodicity at 27.0-days.

Using spectral analysis of low-resolution photospheric 
magnetic field observations from the Wilcox Solar Observatory (WSO)
during solar cycle 21, \inlinecite{Hoek87} 
and \inlinecite{Anto90} found asymmetrical rotation periods 
of 26.9 and 28.1 days for the northern and southern hemispheres
respectively. Combining solar flare observations by hemisphere, asymmetrical 
rotation rates are also detected for prominent flare activity regions, with 
periods of 26.72 and 27.99 days for the northern and southern hemispheres 
respectively (\opencite{Bai90}). Using solar wind velocity data from 1964 
through 1975, \inlinecite{Gos77} reported a pronounced modulation at
27.025 days. Recent work by \inlinecite{Neu00} 
using over three decades of solar magnetic field and solar wind 
data, demonstrated a very persistent signal 
with a well-defined period of $27.03 \pm 0.02$ days. They used
the difference between the maximum and minimum amplitudes of time 
averaged values, binned in solar longitude for a band of rotational 
periods, to select coherent signals. The 
longevity of the signal over several solar cycles suggests a fixed 
longitudinal magnetic structure in the solar interior, possibly a
non-axisymmetric component of the solar magnetic 
field (e.g. \opencite{Stix74}; \opencite{Bal83}; \opencite{Ruz01}).
Analysis of the lowest-order non-axisymmatic modes of the solar magnetic
field using WSO synoptic maps also resulted in the detection of a
27.03-day periodicity (\opencite{Ruz01}). Assuming the observed 
periodicity relates to a 
distinct solar rotation rate does not aid in constraining the signal 
source to a unique solar latitude or interior depth. The equatorial 
rotation rate periods for magnetic tracers have a range of values 
from 26.4 days for young sunspots (\opencite{Zap91}) to 27.1 days 
for large area sunspots (\opencite{How84}). For a thorough review 
of solar differential rotation measurements see \inlinecite{Beck99}.

Previous studies of photospheric synoptic maps have revealed that 
active regions are not distributed randomly, but tend to 
form ``complexes of activity'' (e.g. \opencite{Bum65}; 
\opencite{Gaiz83}; \opencite{Bro90}). In addition, the longitudinal
distribution of major solar flares has been shown to concentrate
in active zones (\opencite{Bai87}). Recently, \inlinecite{Toma00} 
used NSO Kitt Peak Vacuum Telescope (KPVT) magnetic synoptic maps to 
argue that observed persistent bands of active ``nests'' supports a 
stable pattern for sources of flux emerging in the tachocline. They 
determined that these active nests can persist up to seven solar 
rotations. In our study, we investigate possible signal 
sources that last for more than two hundred solar rotations. 
To address the nature of the 27.03-day period signal, and 
the possibility of a fixed or quasi-stable longitudinal 
solar structure, the phase coherency of solar surface 
magnetic activity is investigated using KPVT synoptic data. 
In particular, we utilize the phase variance of signals 
from integrated full-disk solar images as a measure of coherence. 
To localize the solar surface source of the signal, we use spectral 
analysis of temporally sampled KPVT magnetic synoptic 
maps. Additionally, to better understand the longevity of 27.0-day 
period signal, the power and phase coherence are analyzed of the 
international sunspot number time series for the past 120 years.
\begin{figure}
\resizebox{11.5cm}{9.0cm}{\includegraphics{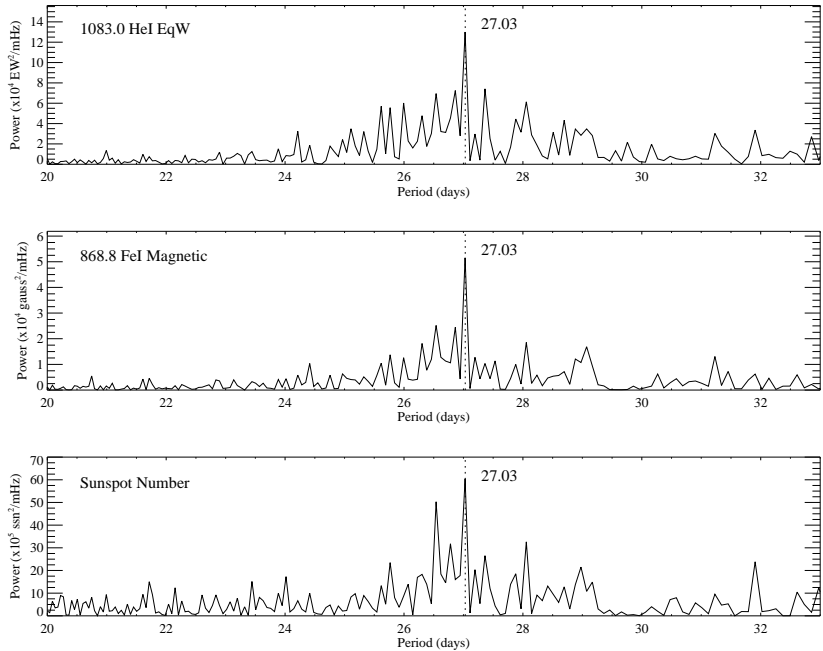}}
\caption{Comparison of power spectra between the KPVT 1083.0 nm 
He I equivalent width (top), 868.8 nm Fe I unsigned magnetic 
flux (middle) full-disk integrated time series, 
and the international sunspot number (bottom) time series. Each time 
series is sampled at a cadence of 1 day and for the same 8839 day 
interval (1977.2 to 2001.4, in units of fractional years). The vertical 
dotted line delineates the 27.03-day period position.}
\label{fig-mPS}
\end{figure}

\section{Integrated Full-Disk Time Series Analysis}

The KPVT integrated full-disk time series used to create the 
power spectra shown in Figure~{\ref{fig-mPS}} are 
disk averages of 1 arc-second pixel solar images of the 
line-of-sight component of unsigned magnetic flux and 
the line equivalent width derived from the 868.8 nm Fe I 
and 1083.0 nm He I lines respectively. At a cadence of one day, 
the signal coverage for the magnetic and the equivalent width time 
series are approximately 70\% and 61\% respectively for the 
8839 day interval, from 1977.2 to 2001.4 (fractional years), used 
to create the spectra in Figure~{\ref{fig-mPS}}. After the temporal 
gaps are filled with a high-order autoregressive 
model (e.g. \opencite{Pro96}), each time series is temporally 
filtered using a non-recursive 
digital filter with a band pass with periods between 
10 and 60 days. We found that the time series gap filling 
has a negligible effect on the final calculated power spectra 
and is used here only to simplify the temporal filtering.

Besides the power spectra for the magnetic and equivalent 
width time series, the power spectrum of the international sunspot 
number, from http://sidc.oma.be, for the same period is 
also included in Figure~{\ref{fig-mPS}}. The three power spectra have 
similar features, including a significant band of power 
between 25.0 to 29.5 days. Previously observed signals at periods 
of 28- and 29-days (e.g. \opencite{Sval75}; \opencite{Shee86}) 
are clearly evident in the three spectra, most especially in the 
magnetic data. Also, even though the signal-to-background ratio of 
the sunspot number spectrum is lower relative to the other two time 
series, the 27.03-day period signal
is notably evident in each of the three spectra. Even with 
the low spectral resolution, the strength and narrowness of the power 
implies a coherent signal. 
However, a large amplitude signal with moderate phase
coherence can result in a higher power spectrum signal relative to 
a smaller amplitude signal with higher coherence (see example in 
the appendix). To better understand the nature of the 27.03-day period
signal coherency, the phase variance of the signal is analyzed in 
the following subsection.
\begin{figure}
  \resizebox{11.5cm}{11.0cm}{\includegraphics{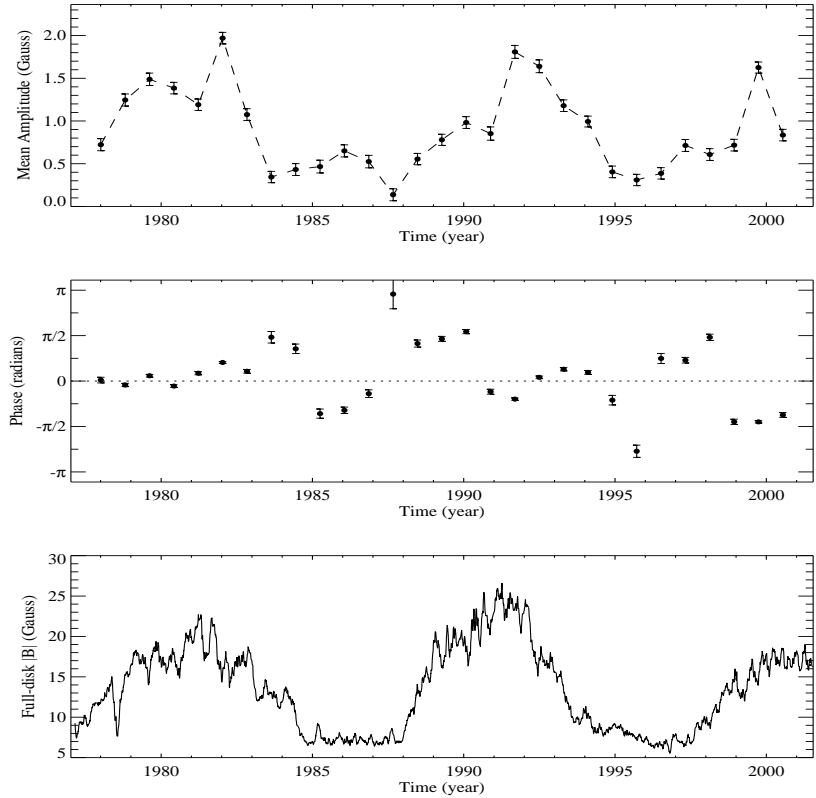}}
  \caption{The estimated mean amplitude (top) and zero-mean
phase (middle) of the unsigned integrated full-disk magnetic field 
time series. The amplitude and phase values are for 29 segments 
with 50\% overlap over the same 8839 day interval as in Figure~1. 
The values are from a non-linear least-squares fit of a cosine 
function to each segment using a fixed frequency with a period 
of 27.03 days. In addition, the ``raw'' integrated 
full-disk field strength, smoothed with a 27-day window, is
displayed for the same period (bottom).
The superimposed error bars on the mean amplitude 
and zero-mean phase values are the 1-$\sigma$ fitting errors.}
\label{fig-phase}
\end{figure}

\subsection{Signal Coherency}

A fixed frequency harmonic signal is defined to be coherent for 
an interval of a time series investigated if the signal has a 
negligible temporal variance in phase. One method to estimate a 
signal's phase is to calculate the instantaneous phase and frequency 
for each time step over the length of a time series using the 
analytical signal from the Hilbert transform (e.g. \opencite{Brace85}). 
However, the 27.0-day signal observed in the three
time series shown in Figure~{\ref{fig-mPS}} is in a crowded 
region of the power spectrum, and the Hilbert transform 
works best with an isolated narrow band signal. Additionally,
wavelet analysis may also give confusing results in 
crowded spectra.

Instead of using a Hilbert transform type analysis, we estimate 
the phase and amplitude of a signal from 
a non-linear least-squares fit of a fixed frequency cosine 
function for overlapping segments of a time series. 
As with the power spectra analyses, the time series are
gap filled with a high-order autoregressive 
model and temporally filtered using a non-recursive 
digital filter with a band pass with periods between 
10 and 60 days. No further processing is done to the data
before or after the time series is segmented. The time steps
that contain gap filled values are given weights of zero during 
the non-linear least-squares fit of each segment.

In Figure~{\ref{fig-phase}}, the temporal variation of
phase and mean amplitude for the same 8839 day interval as shown
in Figure~{\ref{fig-mPS}}, using 29 segments of 589-day duration
with 50\% overlap, are depicted for the KPVT integrated full-disk  
magnetic field strength time series using a fixed frequency with 
a period of 27.03 days. The phase is displayed relative to the 
error weighted mean phase, where the errors are from the fit. 
The signal is mostly coherent during periods of large amplitude.
Note, however that the signal, though coherent, is approximately 
90 degrees out of phase relative to the error weighted mean phase 
for the maximum period from year 1999.5 to 2001. As expected, the
signal mean amplitude extremes correspond reasonably well
with high amplitudes of the 
integrated full-disk magnetic time series shown in 
Figure~{\ref{fig-phase}}. For this 24 year interval the 27.03-day 
period signal phase variance is minimum, i.e. more coherent, 
during maximum solar magnetic activity. Also, note that the signal's
phase wanders approximately $\pm$90 degrees during lower activity
spans. The question of how the phase variance of this signal compares 
to other frequencies is addressed in the following subsection.

\subsubsection{Phase Coherence Spectrum}

As a measure of a signal's coherency, relative to
neighboring frequencies, the inverse phase variance of a 
segmented time series for fixed frequencies within a 
selected frequency window is used to create 
a ``phase coherence spectrum'' (PCS). The basic 
premise of the PCS is that long-lived coherent signals will 
have a smaller phase variance than shorter 
lived or less coherent signals relative to the time
series duration investigated. To create the PCS, the 
phase and amplitude are estimated from a non-linear 
least-squares fit of a fixed frequency cosine function to each 
time series segment. Next, the temporal variation of the phase 
and amplitude are quantified for a discrete frequency 
and segment length. The phase coherence is defined here
to be the inverse phase variance, where the mean phase and the 
phase variance are weighted by the inverse of the phase 
error determined from the non-linear least squares fit. 
The chosen segment length will affect the variance of 
the estimated phase values. As the segment length increases
the temporal phase discontinuities approach the value of 
the mean phase for the time series interval. The nature and 
attributes of the PCS, along with the effects of varying the segment 
length, are explored in more detail in the appendix. 
\begin{figure}[htp]
\resizebox{11.5cm}{8.0cm}{\includegraphics{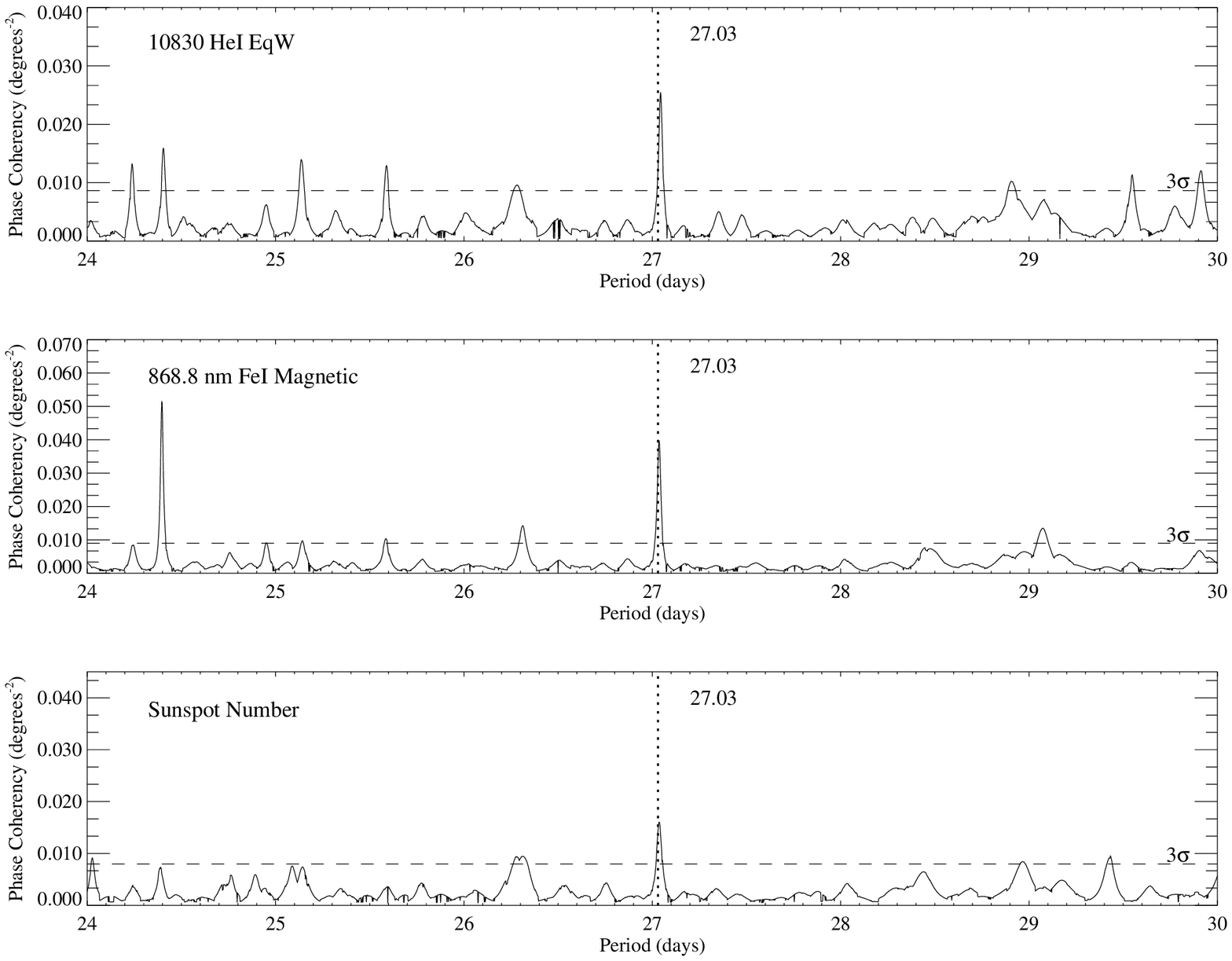}}
\resizebox{11.5cm}{8.0cm}{\includegraphics{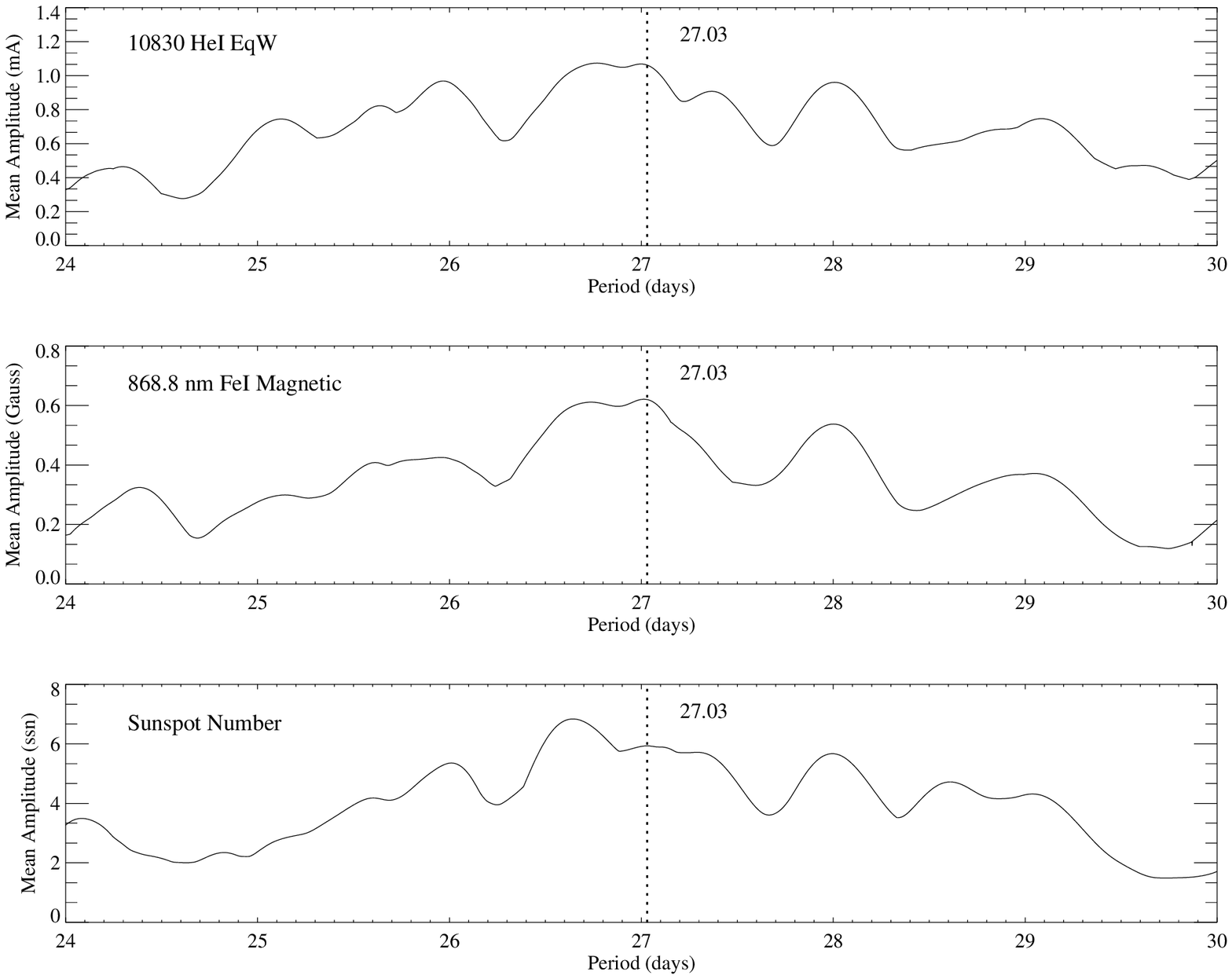}}
\caption{Phase coherence spectra (top three) and  
the error weighted mean amplitude spectra (bottom three) for the 
KPVT integrated full-disk 1083.0~nm He I equivalent width, 
868.8~nm Fe I unsigned magnetic flux time series, and the 
sunspot number time series for the same 8839 day interval used 
in Figure 1. The PCS were created using 9 segments with 50\% overlap 
and widths of 1/5 the total time series duration. The vertical dotted 
line delineates the 27.03 day period position. The horizontal dashed 
line indicates the 3-$\sigma$ noise level.}
\label{fig-mPCS}
\end{figure}

The PCS method is similar to the phase dispersion 
minimization (PDM) method (\opencite{Stel78}). The PDM 
method folds a given time series into phase bins for a selected 
period, then the variances of the binned values are compared for
the chosen frequency band to select out those periods that
correspond to periodic signals. However, unlike
the PDM method, the PCS method estimates and utilizes the temporal
variation of phase and amplitude, along with the fitting errors,
for a given frequency. The temporal error information is critical
for analyzing long lived signals that may have ``dormant'' states,
i.e. periods when there is a very small or no signal amplitude
observed relative to other segments of a time series, e.g. solar
magnetic activity (see Figure~{\ref{fig-phase}}).

The phase coherence and mean amplitude spectra for the magnetic, 
equivalent width and sunspot number time series are shown in 
Figure~{\ref{fig-mPCS}}. All three spectra are from the same 8839 day 
interval as in Figures~{\ref{fig-mPS}} and {\ref{fig-phase}}. 
An example of the usefulness of the phase
coherence and mean amplitude spectra is illustrated with the
28.1-day and 29.0-day period signals which are evident in the power 
spectra in Figure~{\ref{fig-mPS}}. From Figure~{\ref{fig-mPCS}},
the lack of phase coherence and the increased mean amplitude at a 
period of 28 days implies that the signal is primarily 
the result of strong magnetic field regions incoherently 
distributed relative to a weaker but more spatially coherent 
source of the 29.0-day signal.

As with the power spectra in Figure~{\ref{fig-mPS}}, the 27.03-day 
period signal is notable in each of the phase coherence 
spectra shown in Figure~{\ref{fig-mPCS}}. Within the frequency band
between 25 and 29 days, the 27.03-day period signal is clearly
the most coherent.  At this stage, we confirm the results 
of \inlinecite{Neu00}. The highest coherence peak within the magnetic 
PCS is at a period of 24.4 days. This coherence spike is most likely a 
spurious signal because it is not equally coherent in the sunspot 
and 1083 PCS relative to the 27.03-day period signal. Another 
signal of interest, albeit marginally detected, is at a period 
of 26.3 days. This period of rotation is faster than any measured 
magnetic surface tracer. However, this signal may be related to
young sunspots which have a measured rotational 
period of 26.37 days at the equator (\opencite{Zap91}).
Additionally, the maximum mean amplitude within the 24 to 30 day 
period band is near or at a period of 27.0 days in each of the mean 
amplitude spectra. With clear evidence for a coherent solar surface 
magnetic signal in Figures~{\ref{fig-mPS}} and~{\ref{fig-mPCS}}, the 
next step is to better understand the character of the signal source.
In the following section, the solar surface is 
investigated for regions with coherently reoccurring magnetic activity 
during the past two solar cycles that may account for the 27.03-day 
period signal.

\section{Spatially Resolved Time Series Analysis}

\begin{figure}
\resizebox{11.5cm}{6.0cm}{\includegraphics{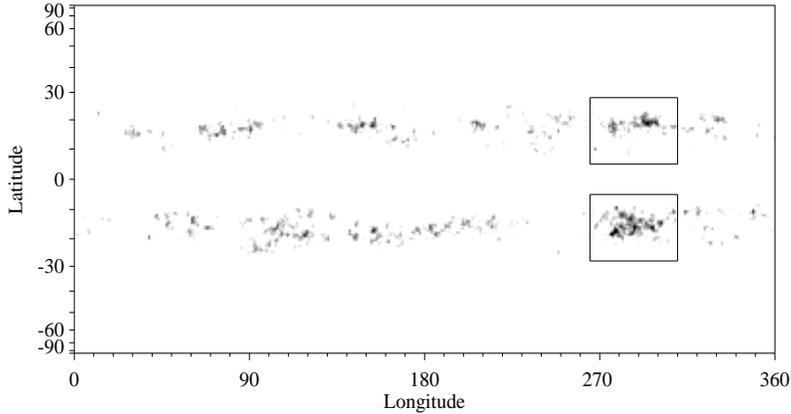}}
\caption{The average of 330 unsigned magnetic flux synoptic 
maps using a rotation period of $27.03\pm.02$ days. The maps
are from KPVT observations during years 1977.70 through 2001.34
(Carrington rotations 1645 through 1974). Regions
between white and black are between 2-$\sigma$ and 4-$\sigma$ 
above the mean respectively. Longitude values are relative to 
Carrington rotation 1645. The two primary complexes of 
magnetic activity are highlighted by the boxed regions.}
\label{fig-hotmap}
\end{figure}

\begin{figure}
\resizebox{11.5cm}{18.0cm}{\includegraphics{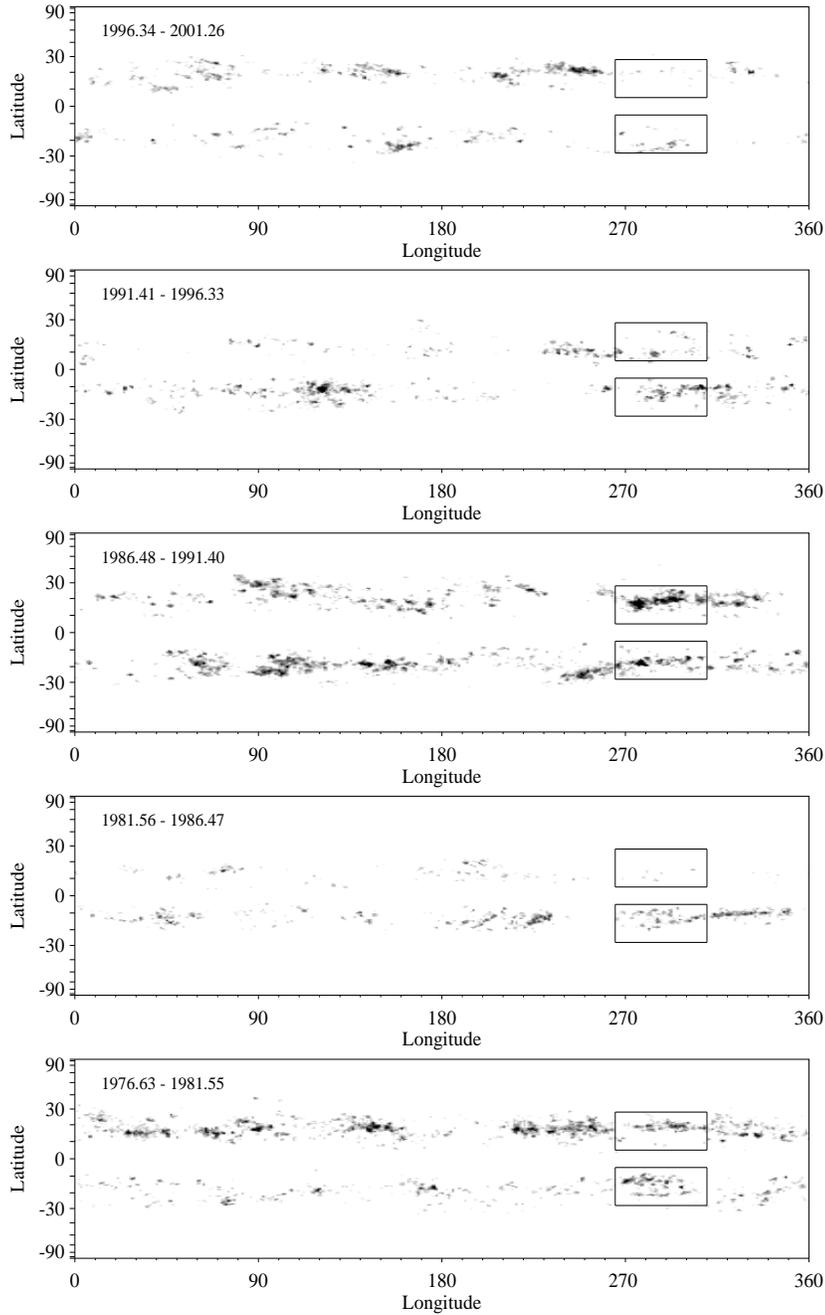}}
\caption{Averages of sets of 66 unsigned magnetic flux synoptic 
maps using a rotation period of $27.03\pm.02$ days. Regions
between white and black are between 2-$\sigma$ and 6-$\sigma$ 
above the average mean of the five averaged maps respectively. 
Longitude values are relative to Carrington rotation 1645.
The two boxed regions in each image corresponds to the highlighted 
regions in Figure~{\ref{fig-hotmap}}.}
\label{fig-mhotmaps}
\end{figure}

To explore if any particular solar surface region is a probable 
source of the 27.03-day signal, nearly 24 years of the KPVT 
unsigned photospheric magnetic flux synoptic maps are averaged 
using a rotational period of 27.03 days. The KPVT synoptic 
maps used here are described by \inlinecite{Harvey80}, and also 
see \inlinecite{Toma00}. Shifting each synoptic map relative to 
the previous maps using 
a rotation period of $27.03\pm.02$ days, reveals two long-lived 
complexes of activity exhibited in Figure~{\ref{fig-hotmap}}. 
The two stronger-than-average regions of magnetic
activity (boxed areas) are nearly aligned at 290 degrees longitude, 
one on each hemisphere, in the Carrington Rotation
number 1645 reference frame. The alignment in longitude
of the two complexes of activity is suggestive of a fixed
magnetic sector structure. However, when the synoptic maps are 
averaged over shorter periods, shown in Figure~{\ref{fig-mhotmaps}}, 
the boxed regions do not reveal persistent activity relative 
to other longitudes. The two boxed regions 
highlighted in Figure~{\ref{fig-hotmap}} are
primarily the result of strong activity during solar cycle 21
(between 1986 and 1996), displayed in Figure~{\ref{fig-mhotmaps}}.

\begin{figure}
\resizebox{11.5cm}{18.0cm}{\includegraphics{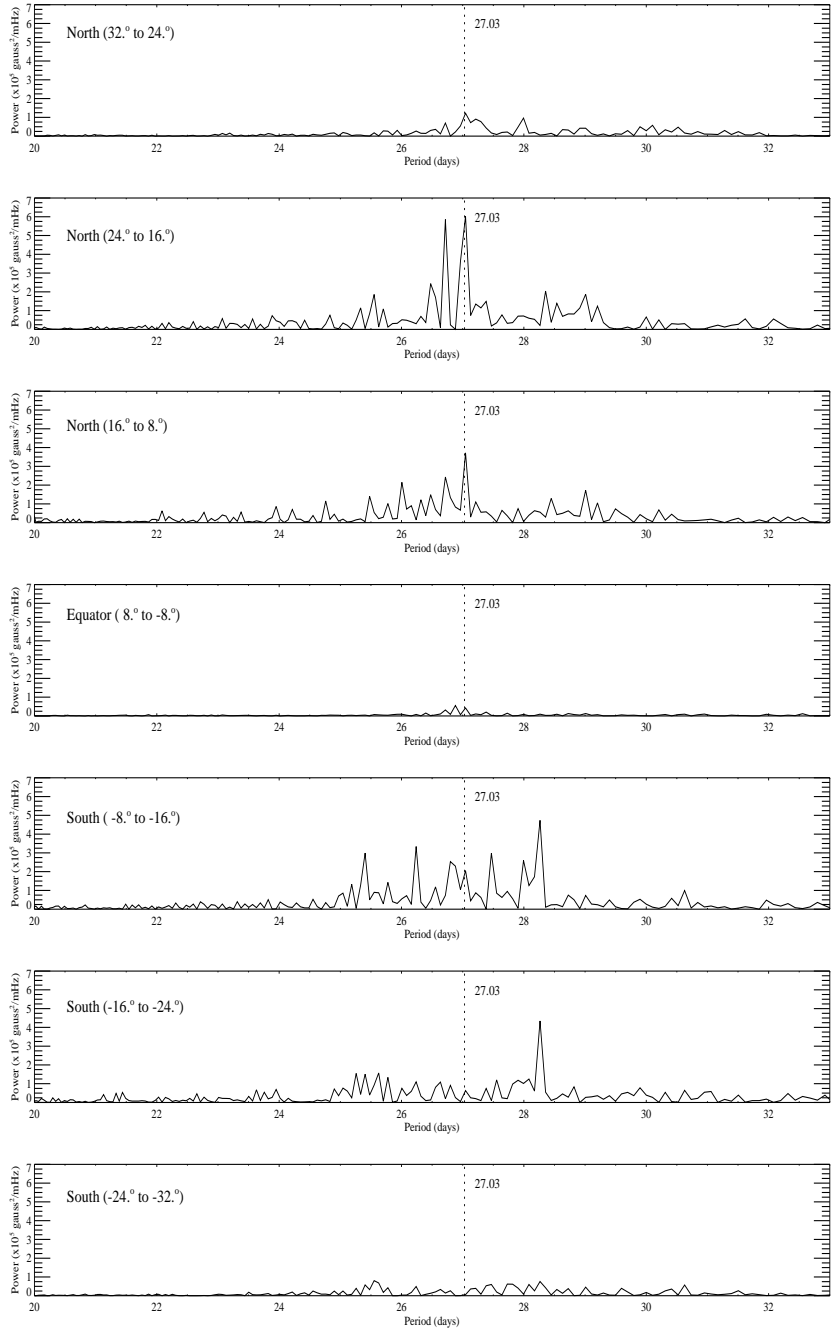}}
\caption{Comparison of power spectra for northern (top three),
equatorial (center), and southern (bottom three) latitude bands from 
330 unsigned magnetic synoptic maps for the same period used to 
create the average synoptic maps in 
Figures~{\ref{fig-hotmap}} and~{\ref{fig-mhotmaps}}. The vertical 
dotted line delineates the 27.03 day period position.}
\label{fig-latPS}
\end{figure}

In addition to averaging synoptic maps with a 27.03-day
rotational period, the synoptic maps were temporally sampled
to create latitudinal time series for spectral analysis. 
For this analysis the unsigned magnetic flux synoptic maps 
are binned in latitude and longitude. 
As with the integrated full-disk spectral analysis, temporal gaps 
are filled with a high-order autoregressive model, and each 
time series is temporally filtered using a non-recursive digital filter 
with a band pass between 10 and 60 days. The power spectra for middle 
to low latitudes are displayed in Figure~{\ref{fig-latPS}}. 
As expected from long-term sunspot observations, the 
lower and higher latitudes contribute negligible power to the 
integrated full-disk power spectrum.

From Figure~{\ref{fig-latPS}}, it is evident that the 27.03-day 
period signal is distinctly from the northern hemisphere. 
Additionally, the $\sim$28-day signal is predominantly 
from the southern hemisphere. These results support the previous
findings of asymmetric rotation periods for the solar hemispheres,
$26.9 \pm 0.2$ and $28.1 \pm 0.2$ days for the northern and southern 
respectively, reported by \inlinecite{Hoek87} 
and \inlinecite{Anto90}. From Figure~{\ref{fig-hotmap}}, the 
maximum amplitude in the northern hemisphere complex is at a 
latitude of $18 \pm 2$ degrees. This result agrees with the 
finding by \inlinecite{Anto90} that the 26.9-day period signal 
source is centered at a latitude of 15 degrees north, covering a 
latitude zone of approximately 24 degrees wide. At a latitude of 18 degrees, 
the 27.03-day period signal is consistent with the solar interior rotation 
rate of 27.06 days at a depth of $0.95$ \rsun, obtained from helioseismic 
studies using data from the Michelson Doppler Imager (\opencite{Schou98}).

\section{Signal Longevity}

To estimate the longevity of the 27.03-day period signal, we analyzed 
more than 120 years of the international sunspot number time series. 
Previously, \inlinecite{Bog82} found evidence for preferred 
activity longitudes using spectral analysis of the sunspot number 
time series. The length and generally consistent character of the sunspot 
number time series, both in quality and cadence, makes it 
ideal for testing the long-term persistent nature of a particular 
signal. Furthermore, the sunspot number time series is of interest
since its power spectrum shown in Figure~{\ref{fig-mPS}} is 
remarkably similar to the KPVT spectra and therefore appears to
be useful as a proxy for past solar surface magnetic activity. 
\begin{figure}
\resizebox{11.5cm}{13.0cm}{\includegraphics{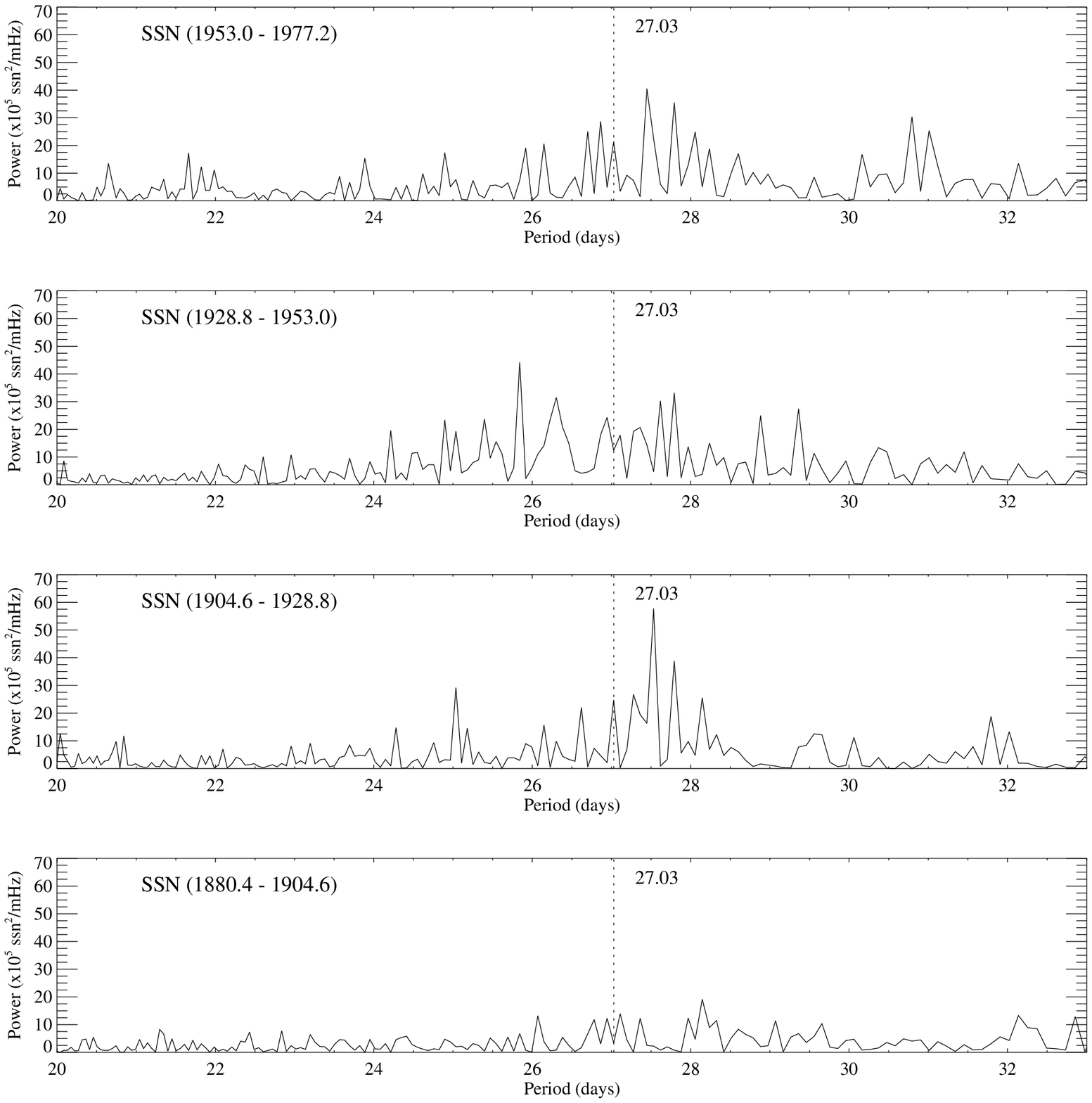}}
\caption{The international sunspot number (SSN) power spectra 
for four consecutive periods during the past 120 years. 
The time series cadence of 1-day and interval length of 8839-days
used for each power spectrum is equivalent to those of the 
comparison in Figure~\ref{fig-mPS}. The interval values for each time 
series (in parentheses) are in fractional years. The vertical dotted 
line delineates the 27.03 day period position.}
\label{fig-mSSN}
\end{figure}
\begin{figure}
\resizebox{11.5cm}{13.0cm}{\includegraphics{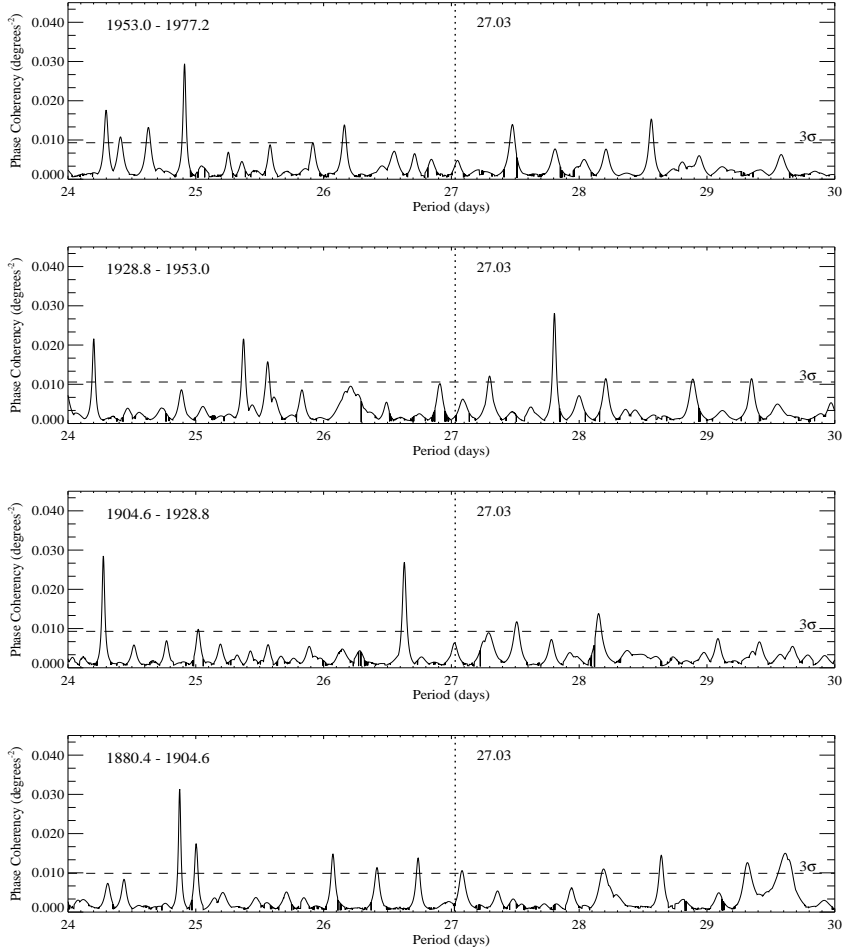}}
\caption{The phase coherence spectra for 
the sunspot number time series over the past 120 years, using 
9 segments with 50\% overlap and widths of 1/5 the total time 
series length. Each time series has the same 1-day cadence 
and interval length of 8839-days as used for those in 
Figure~\ref{fig-mPS}. The vertical dotted line delineates 
the 27.03 day period position. The horizontal dashed 
line indicates the 3-$\sigma$ noise level.}
\label{fig-mpSSN}
\end{figure}
\begin{figure}
\resizebox{11.5cm}{13.0cm}{\includegraphics{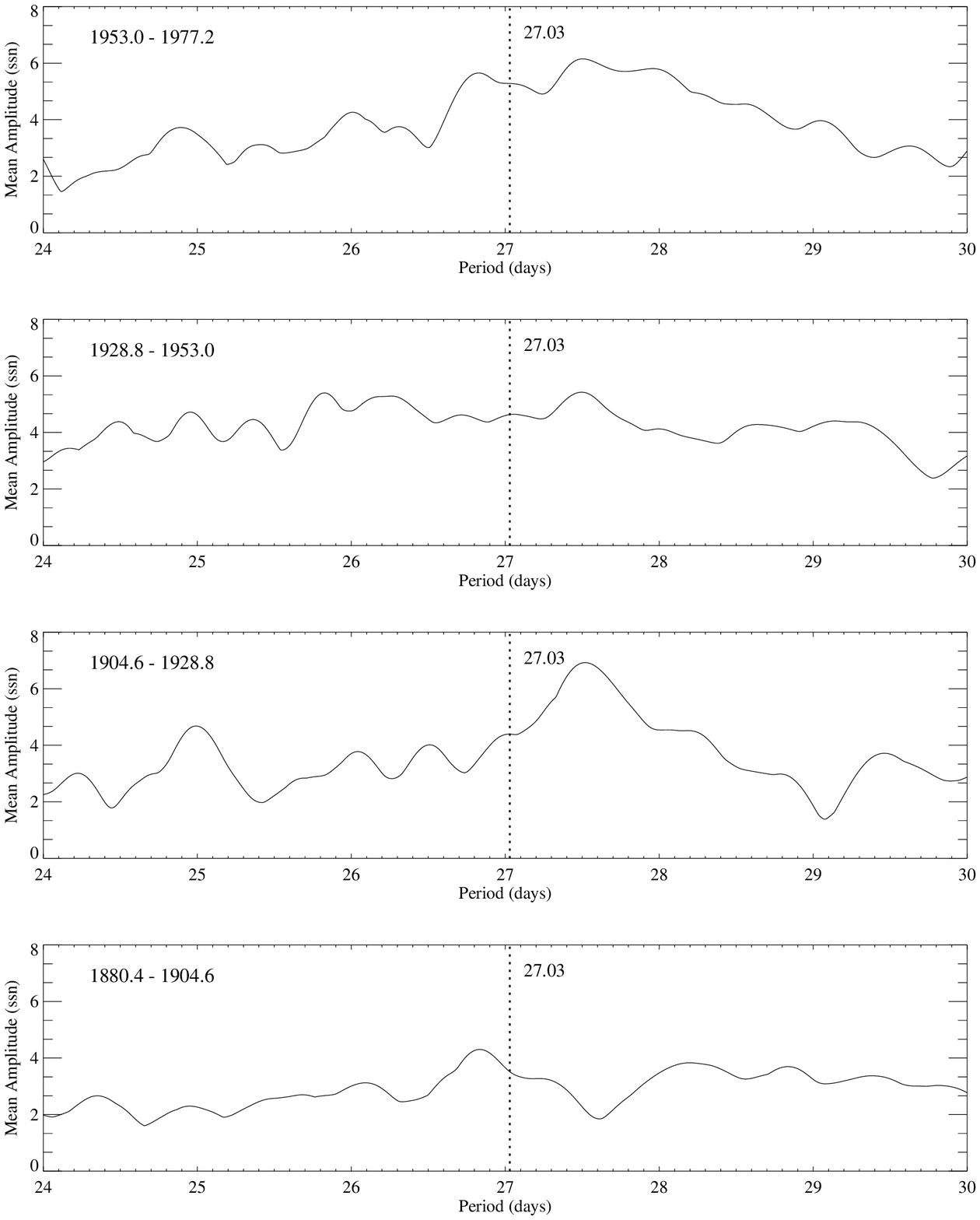}}
\caption{The error weighted mean amplitude for the international
sunspot number time series during the past 120 years, using 9 segments 
with 50\% overlap and widths of 1/5 the total time series span.
Each time series has the same 1-day cadence and interval 
length of 8839-days as used for those in Figure~\ref{fig-mPS}. The 
vertical dotted line delineates the 27.03 day period position.}
\label{fig-maSSN}
\end{figure}

In addition to the sunspot number power spectra for the most 
recent 24 year interval displayed in Figure~{\ref{fig-mPS}}, spectra for 
four additional periods of equal length during the past 
120 years are exhibited in Figure~{\ref{fig-mSSN}}. Using the same 
time series analysis method as discussed in section 2, the 27.03-day 
period signal is not found to be significant in the power spectra
for the years between 1880 and 1977 as compared to 
the 1977 to 2001 spectrum. The only frequency to have notable power
in more than one of the five intervals is at a period of approximately
27.5 days during years 1904 through 1977 (see the top three graphs in
Figure~{\ref{fig-mSSN}}).

The phase coherence and mean amplitude spectra are revealed 
in Figures~{\ref{fig-mpSSN}} and~{\ref{fig-maSSN}} respectively. 
Again, the 27.03-day period signal is found to be
insignificant for the years between 1880 and 1977. 
In Figure~{\ref{fig-mpSSN}}, the 27.5-day period signal has
marginally significant phase coherence during the intervals 1904-1928 and 
1953-1977. Nonetheless, the maximum mean 
amplitude is centered on a period of 27.5 days between years 1904 
and 1977 (top three plots in Figure~{\ref{fig-maSSN}}). This 
corresponds to the persistent $27.5\pm0.5$-day period signal reported 
by \inlinecite{Bog82}.

Other than the 27.5-day period signal, no other frequency is found 
to have notable coherence during the past 120 years. However, 
the signal-to-background ratio (S/B) of the sunspot number power 
spectra in Figure~{\ref{fig-mPS}} is lower than 
that of the magnetic or 1083 spectra. Likewise, the phase coherence
in Figure~{\ref{fig-mPCS}} of the sunspot number time series is 
significantly lower than in the
other two time series. The poor S/B and lack of coherence
may be reflective of the sunspot number time series, i.e. ambiguity 
in the scaling used to calculate the sunspot number value, or
because sunspots last for a shorter time than the plages seen in 
the magnetic and 1083 equivalent width data.

\section{Conclusions}

A strong 27.03-day period signal, initially reported 
by \inlinecite{Neu00}, is clearly detected in over two decades of 
full-disk KPVT photospheric magnetic flux synoptic data. Using phase 
coherence analysis, the coherency of the  27.03-day signal is found to be 
significant, relative to neighboring periods, for the past two decades. 
The signal is notably coherent during periods when the signal
amplitude is large for the past 24 year term. 
From analysis of spatially-resolved KPVT magnetic flux synoptic maps, 
the origin of the 27.03-day period signal is found to be long-lived 
complexes of activity in the northern hemisphere at a 
latitude of approximately 18 degrees. 
These quasi-stationary magnetic patterns may be the result of a flux 
transport phenomenon, e.g. the model proposed by \inlinecite{Shee87} 
where the rotational shear of magnetic flux is offset by supergranular 
transport and the meridional flow. Additionally, the marginal pattern
coherence could be produced by organized emergence of flux due to a 
quasi-stable interior structure related to the solar 
dynamo (e.g. \opencite{Ruz98}; \opencite{Ruz00}). A detailed analysis of 
magnetic activity, during the past 24 years within a latitude band of 
a few degrees centered on 18 degrees, using individual magnetograms 
instead of synoptic maps would help clarify the signal source, i.e. 
young sunspots verses small area sunspots.

However, the 27.03-day signal lacks clear coherence in the 
international sunspot number time series for the period 
from 1880 to 1977. The signal with the greatest longevity during most 
of the interval 1904 through 1977 is at a period of 27.5 days. 
This 27.5-day periodicity, though not particularly coherent, corresponds 
to a rotation rate that is typically observed for the latitude bands
associated with sunspot activity. There is no evidence for a 
long-lived (more than three 11-year solar cycles) phase 
coherent signal that would indicate a permanent solar magnetic 
non-axisymmetric component at a fixed solar interior depth. 

\section{Acknowledgments}

This research was supported in part by the Office of Naval Research 
Grant N00014-91-J-1040. The NSO-Kitt Peak data used here 
are produced cooperatively by NSF/AURA, NASA/GSFC, and NOAA/SEC.

\appendix

The ``phase coherence spectrum'' (PCS) is created by estimating 
the phase and amplitude for each segment span of a segmented time series 
at fixed frequencies within a chosen frequency band window. From 
a non-linear least-squares fit to a cosine function 
of the form $f_i(t) = a_i \cos(\omega t + \phi_i)$, the amplitude, 
$a_i$, and phase, $\phi_i$, are estimated for each time series 
segment $i$, where $\omega$ and $t$ are the cyclic frequency and time
index respectively. The phase coherence at a given frequency $\omega$,
over the full length of a time series, is defined here as 
\begin{equation}
  \Phi_{\rm coh}(\omega)= \frac {1} {\sigma^2_{\rm phase}},
\end{equation}
where $\sigma^2_{\rm phase}$ is the error weighted phase variance, 
defined as
\begin{equation}
  \sigma^2_{\rm phase} =  
      \frac {\D\sum_{i=1}^N\lbrack (\phi_i - \bar{\phi})^2 w_i \rbrack}
            {(N-1)\D\sum_{i=1}^N w_i}.
\end{equation}
The total number of segments is represented by $N$. The mean phase 
and phase variance are weighted 
by the inverse of the phase error, $w= 1 / \sigma_\phi^2$, determined 
from the non-linear least squares fit. One key caveat when calculating 
the PCS is to sift the estimated phase values to within 
$\pm \pi$ of the error weighted mean phase of all the segments. This 
is accomplished by
adding $\pm 2\pi$ to each phase value, which by the character of its 
definition is constrained by mod $2\pi$, to minimize the error 
weighted phase variance. However, due to the cyclic nature of the 
phase, the phase time series for a given frequency will have spurious 
discontinuities. An example of this is illustrated in the phase time 
series shown in Figure~{\ref{fig-phase}} (middle graph) near year 1995.

\begin{figure}
\resizebox{11.5cm}{9.0cm}{\includegraphics{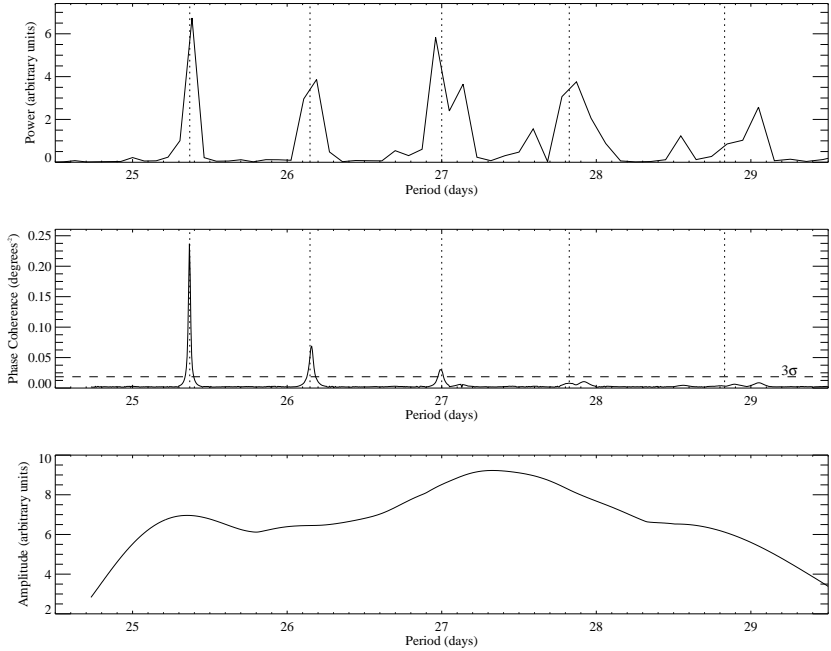}}
\caption{The power (top), phase coherence (middle), 
and the mean amplitude (bottom) spectra for a given realization of
a simulated time series. The time series includes five signals with
coherencies, left to right, 100\%, 80\%, 60\%, 40\%, 20\%, and
unit-less excitation amplitudes of 5, 5, 10, 10, 10  
respectively. The PCS was created using 21 segments with 50\% overlap 
and widths of 1/11 the total time series span. The vertical
dotted lines delineate the frequency of the simulated signal. 
The horizontal dashed line in the phase coherency spectrum (middle) 
indicates the 3-$\sigma$ noise level. }
\label{fig-pcsEX}
\end{figure}

A pure sine or cosine signal that remains in phase over the length of 
a time series, and with no noise, would have a phase coherence value 
(from Equation~1) of infinity. However, this unconstrained upper limit 
is not a problem with a time series that includes noise. Nonetheless,
if a signal of interest is sinusoidal and nearly noise free, the phase
variance values from Equation~2 can be employed instead of the phase
coherence. The phase coherency lower limit is constrained by the 
maximum variance allowed after the phase sifting process. The lower 
limit can be estimated from Equations~1 and~2 by assuming an equal number 
of segments have phases that are either $90$ or $-90$ degrees from 
the mean, with equal error weights for each segment. For a given time
series realization it is possible to have spurious high phase coherent 
signals. Unless clearly associated with features detected in the
power spectrum, these spurious signals are noticeably narrower than
a PCS signal.  One method to improve the signal-to-background ratio of 
the estimated PCS is by averaging the PCS calculated using various 
segmentation lengths.
\begin{figure}
\resizebox{11.5cm}{12.0cm}{\includegraphics{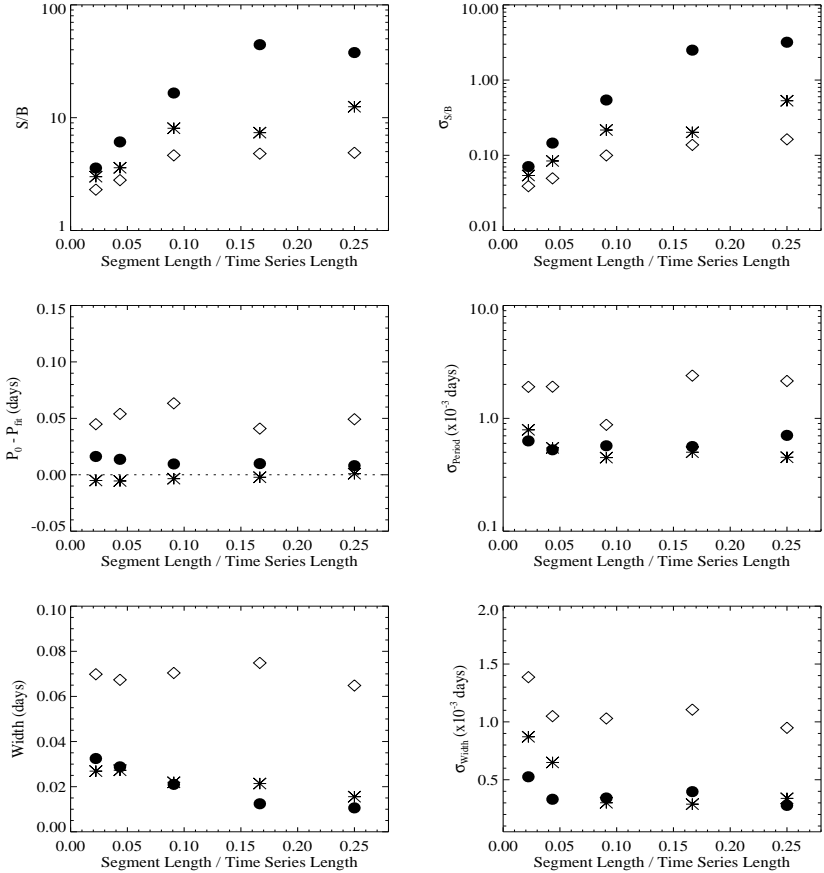}}
\caption{The phase coherence signal-to-background ratio (S/B), residual of 
the difference between the true and the fitted signal 
period, and the fitted Gaussian full-width as a function of segment 
length. The values are from spectra using a simulated
time series with three signals differing only by phase coherence
percentage over the length of the time series: 80\% (solid circles), 
60\% (stars), 40\% (open triangles), and unit-less excitation 
amplitudes of 5, 10, 10 respectively. The left-hand column displays the 
mean S/B, signal period and width estimates of 500 realizations at each 
segmentation length position. The right-hand column depicts the
standard deviations of the distributions of the estimated parameters.
The horizontal dotted line, in the middle left-hand column plot, 
delineates the simulated signal position.}
\label{fig-MC}
\end{figure}

An example power spectrum, phase coherence spectrum, and mean 
amplitude spectrum for a single realization of a simulated time 
series are shown in Figure~{\ref{fig-pcsEX}}. The simulated time 
series is 8250 days in length, with 1 day cadence, and includes 
five signals with coherencies of 100\%, 80\%, 60\%, 40\%, 20\% 
and unit-less excitation amplitudes of 5, 5, 10, 10, 10  
respectively. The coherency percentage is defined here, albeit
arbitrarily, by dividing the simulated time series interval into
five segments, where in-phase segments for a given signal are
excited with a relative phase of zero. The out-of-phase segments 
are excited to have a phase with a mean absolute difference of 
approximately 100 degrees relative to the mean phase of the time series 
interval. The Gaussian white noise added to the 
simulated time series is scaled in amplitude such that 
the 60\% coherent signal has a S/B of 6.4 estimated from the 
distribution of signal fits to the power spectrum.

The power spectrum in Figure~{\ref{fig-pcsEX}} was created using the 
same time series analysis method as discussed in section 2. Note that 
from the power spectrum it is unclear if the 60\% and 40\% coherence
signals are more or less coherent than the 80\% coherence signal, 
whereas the PCS gives some indication of the relative coherence. Also, 
note the narrowness and broadness of the phase coherence and mean 
amplitude signals respectively in Figure~{\ref{fig-pcsEX}}. The narrowness, 
or sensitivity, of phase coherence signal is partly due to the estimation 
of the phase within the cosine function, whereas the amplitude 
signal is broad as a result of being a multiplicative term in the 
non-linear least-squares fit. Though the resolution of the PCS is 
arbitrary, to account for the narrowness of phase coherence signal, a 
factor of ten or greater in frequency resolution than would be 
used in a given frequency band with standard fast Fourier analysis 
is a convenient exploratory resolution.

The phase coherence analysis stability of the simulated time 
series is illustrated by the Monte Carlo simulation results 
shown in Figure~{\ref{fig-MC}}. These PCS results are derived from 
Gaussian fits, for 500 realizations, to the inner three signals with 
coherencies of 80\%, 60\%, and 40\%, illustrated for one realization in 
Figure~{\ref{fig-pcsEX}}. The left-hand column of the figure displays 
the mean signal-to-background ratio (S/B), signal period and 
width estimates of 
500 realizations at each segmentation length position. The right-hand 
column shows a measure of the random errors using the standard 
deviations of the distributions of the estimated parameters. 
As indicated in Figure~{\ref{fig-MC}}, the PCS signal-to-background ratio
decreases with shorter segment lengths due to fewer cycles of the 
given period being fitted, in addition to the growing influence 
of the shorter lived coherent signals and the background. 
A segmentation overlap of 50\% is used in all of the time series PCS 
analysis done in this paper. With overlapping segments, the temporal 
resolution is improved while preserving the signal sampling used for 
the non-linear least-squares fit. 
Also highlighted in Figure~{\ref{fig-MC}} are the limitations of detecting 
signals that are marginally coherent over an interval analyzed. The S/B 
of the 40\% coherence signal is notably lower in value relative to the 60\% 
and 80\% coherence signal values. Also, the measured frequency position 
for the 40\% coherence signal is systematically offset from the true 
value, along with a broader fitted signal width. Nonetheless, the 
Monte Carlo results
demonstrate that the phase coherence spectrum is a beneficial complement 
to power spectrum analysis as a measure of 
the relative phase coherence for signals under investigation.


\end{article}

\end{document}